# The Fellowship of LIBOR: A Study of Spurious Interbank Correlations by the Method of Wigner-Ville Function

**Peter Lerner**[1]

## Abstract

The method of the Wigner-Ville function proposed by Wigner, (1932) and Ville (1947) is widely used in quantum statistical mechanics and signal processing and historically preceded the continuous-time wavelets. (Gabor, 1946) Here it is proposed for the studies of the financial time series. One of the advantages of the Wigner-Ville function is the possibility to easily visualize the results and use image analysis software to analyze the time series, especially pertinent in the modern era of "big data." In the current paper, we use the Wigner-Ville function for the "toy" problem of the suspected manipulation of the LIBOR quotes by the member banks.

[1] Wenzhou-Kean Business School (Wenzhou, China), pblerner@syr.edu and plerner@kean.edu.
The author extends deep thanks to the participants of the seminar at the business school of Shanghai Maritime Technical University, especially J. J. Mi, World Finance Conference section chairman Sergio Puente and my indomitable discussant Damiano Bruno Silipo, as well as participants of the Vietnam Symposium on Banking and Finance, Christoph Wegener—the chair of my session and R. Erbe, a thorough and attentive reviewer. Special Section 4 was added in response to the comments at the SMTU seminar and by D. B. Silipo. Author is responsible for all the errors.

## 1 - Introduction

In the current era of "big data", the methods borrowed from physics and signal processing found a broad area of implementation in the field of financial data analysis. Particularly popular is the method of wavelets, rediscovered several times—the continuous history probably begins with Gabor in 1946. (Mallat, 1999, Percival, 2000) Yet, an even older method of Wigner-Ville function (WVF) invented by Wigner in 1932 (Wigner, 1932) and applied by Ville for the time series (Ville, 1947) did not became popular in finance. In particular, relative lack of the popularity of the WVF method might be attributed to the fact that, unlike the wavelets, the WVF is a sesquilinear function of its argument (the stochastic process in question). Yet, the possibility to study a stochastic process simultaneously in the time and frequency domain, especially useful in the modern era of the "big data" and easily visualize the results can give a boost of popularity to this method for the study of the financial time series.

Phase-space methods have been used for a long time in physical acoustics, in particular, in speech recognition (see, e.g. Cohen, 1995). A characteristic picture why these method can provide information, which cannot be provided separate by time or frequency analysis is given in Fig. 1 (adapted from Myers, 2017).

This paper deals with several data arrays of about ~1,000,000 data points, which cannot be characterized as "big data", now consisting of terabyte-scale databases. Yet, this toy problem analyzing the residuals from submission of credit quotes by the banks, which were the members of the "LIBOR fellowship" during the period in question, can demonstrate both the successes and difficulties of the implementation of the Wigner-Ville function, as well as other quasi-distributions (Hillery, O'Connell, Wigner and Scully, 1984).

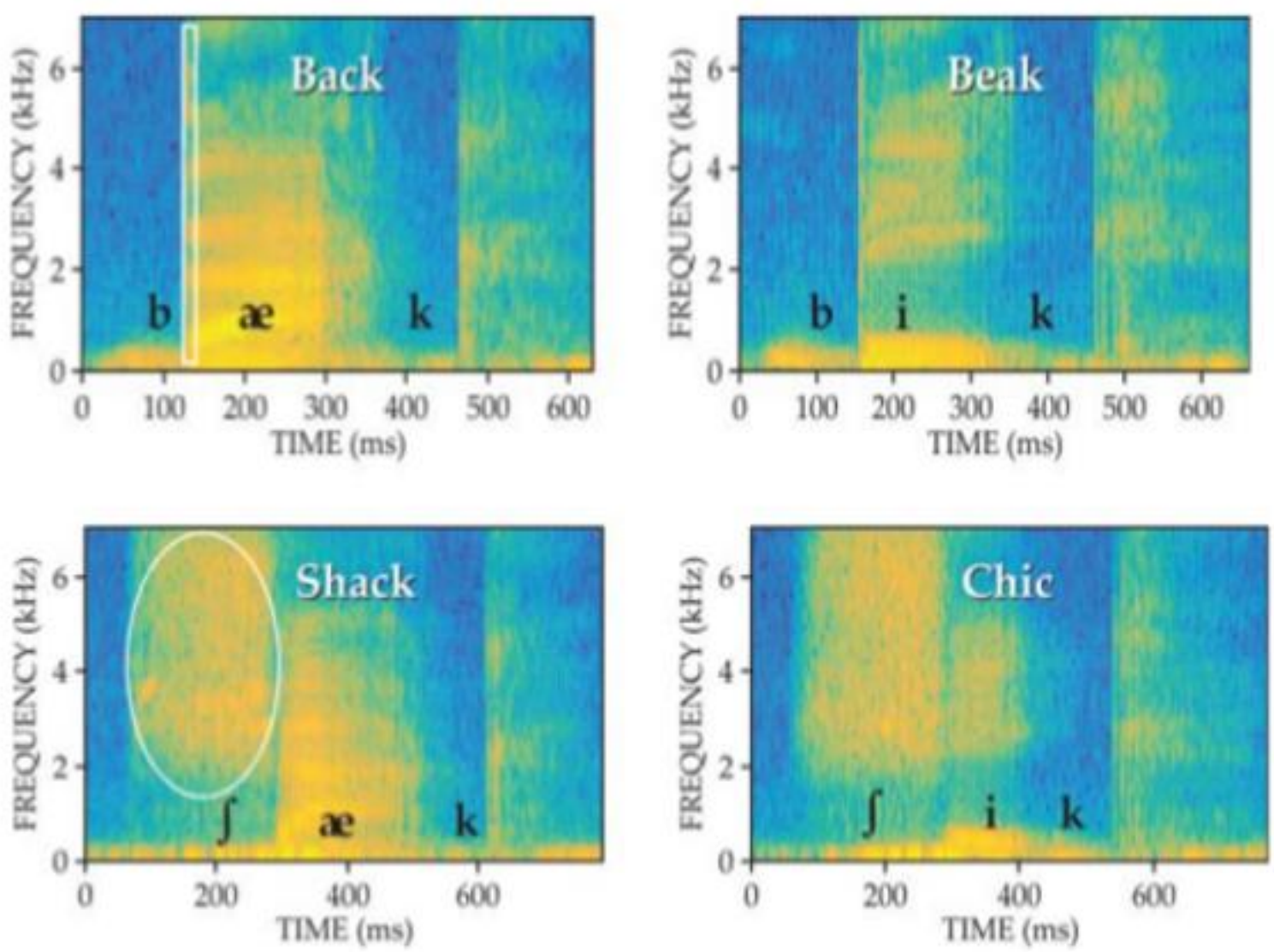

Fig. 1 Speech recognition based on the time-frequency analysis. In the upper set of pictures, the words are being distinguished by the time interval; in the lower set of pictures—by the frequency spectra (adapted from Myers, 2017).

The residuals of the de-trended regressions are studied using the Wigner-Ville function, which is widely used in physics and has been applied for image recognition. The example of LIBOR time series filtering is used to advertise a new and powerful method, which can be applied in other fields of finance. The power of the proposed method is that it allows the analysis of the entire ($\omega$, t) phase space of the time series (see the next section). While the attempts to use the phase space for econometric analysis have been undertaken in a completely different context (e.g. Wei and Leuthold (2012), Smug, Ashwin and Sornette (2017)), the application of the Wigner function to the analysis of the financial time series is new.

While WVF methods are similar to the wavelet approach, the advantages of the WVF methods with respect to wavelets are several-fold. First, unlike the wavelet representations, which do not have an obvious dynamic equation associated with them, the WVF has a complicated, but still existing equation for phase-space dynamics (see Hillery, McConnell, Scully and Wigner, 1984, as well as Appendix B). Second, we can apply well-known algorithms invented for the visual image processing to the WVF maps (see Cichocki, 2002 and Parker, 2010 for more details). Third, unlike the wavelet approach, which has an arbitrary choice of kernel and where there is always a possibility that the results depend on the choice of wavelets, in WVF approach, the time-shifted series itself serve as a kernel. The intuition behind the last statement is that only systematic patterns survive the time-shifting while the randomness gets averaged out on computing mathematical expectations. Fourth, finally, the Wigner function has a complicated, yet a dynamic evolution equation (see Appendix A).

The only limitation specific of the Wigner function method known to the author is its impossibility to verify causality in the spirit of Granger. (Berzuini, 2012) This happens because methods based on Fourier transform are symmetric with respect to the time inversion. To distinguish between past and future, one might use Laplace transform. Possible approach is outlined in Appendix B.

Mathematically, we test possible clustering patterns in the correlation function (Hong, 2006) between residuals in both the time and the spectral domains:

$$C_{ik}[\tau] = E_t[e_i(t - \tau/2) \cdot e_k(t + \tau/2)] \tag{1}$$

where *i, k*÷1, 18 and τ—is the time shift. The patterns are detected by the method of aliasing, which is widely used in image-detection technology (Mallat, 1999). In the language of statistics, Equation (1) represents the family of the covariance matrices of the time-shifted time series.

## 2 - A Primer on the Wigner-Ville Function

The Wigner function was introduced by Eugene Wigner (1932) as an attempt to introduce a concept of phase space in quantum mechanics. It was tangential to the developments in contemporary quantum mechanics but was revived post-war by Moyal (1947) and Ville with respect to the time series (1952). The Wigner function comes in two flavors: the original and the Wigner-Ville modification for the time series. It is unclear at this point which one has any application (if any) in financial mathematics. The original Wigner formulation presumes the phase space of a Hamiltonian dynamical system of dimension 2*N* (symplectic manifold) where *N* is a number of the state variables, the points of which are numbered by two variables: $(\vec{p}, \vec{q})$. In reference to the mechanical origin of the construction, the state variables $\{q_i\}$ are identified with coordinates and $\{p_i\}$ with momenta. In this formulation, time is a parameter. Then, if we have a distribution of the state variable of the dynamical system $f(\vec{q})$, the Wigner function is defined by the following Fourier integral:

$$W(\vec{p}, \vec{q}) = \int f\left(\vec{q} + {}^{\vec{q}'}/_{2}, t\right) \cdot f^*\left(\vec{q} - {}^{\vec{q}'}/_{2}, t\right) e^{-i\sum_{i=1}^{N} p_i \cdot q_i'}\, d^N q' \qquad (2)$$

where it is defined as a bilinear integral over the distribution *f(q)*. In the context of the statistics, this distribution is always real, but we retain the complex conjugation sign to preserve the symmetry of the Wigner function definition with the distribution for the “conjugate” state variables *f(p)*.

The formal definition of the WVF for the time series is:

$$W[\tau, \omega] = \int_{-\infty}^{\infty} x\left(\tau - {}^{t}/_{2}\right) \cdot x^*\left(\tau + {}^{t}/_{2}\right) e^{-i\omega t} dt \qquad (3)$$

In Equation (1), *x(t)* is a signal (time series) and star means, as usual, the operation of complex conjugation. Some applications of the WVF to the analysis of images are described in Mallat (1999). The advantage of WVF is that it simultaneously measures time and frequency domains.

If we deal with a vector-valued stochastic process, the natural generalization of the definition of the Equation (2) is as follows. In particular, if we consider the Wigner transform of the vector components of the same function $\{f_i\}$, then the matrix elements will be expressed by the equation:

$$W_{ik}[\omega, \tau] \equiv W\{f_i, f_k\} = \int f_i\, (t - \tfrac{\tau}{2}) f_k^*\left(t + \tfrac{\tau}{2}\right) . e^{-i\omega t} dt \qquad (4)$$

In our application, matrix indexes $i$ and $k$ refer to the different instruments in the portfolio and the functional indexes—to the time measured on the daily basis.

The Wigner function has some important properties, which are discussed below according to (Hillery, O'Connell, Scully and Wigner, 1984). The integration of the Wigner function over one argument produces the modulus square of the distribution function of the other argument:

$$\int W(p,q)dp = |f(q)|^2 \tag{5}$$

$$\int W(p,q)dq = |f(p)|^2$$

This property applied to the WVF integrated over time that it provides a <u>periodogram</u> of frequency (used, e.g. in identification of fat tails in econometrics) and the frequency integral provides conventional <u>correlogram</u> of the time series:

$$C_{ik}[\tau] = E\left[x_i\left(\tau - {}^{t}/_{2}\right) \cdot x^{*}{}_{k}\left(\tau + {}^{t}/_{2}\right)\right] = \int W_{ik}[\tau,\omega]d\omega \tag{6}$$

$$\boldsymbol{I}_{ik}[\tau] = E\left[\tilde{x}_i\left(\omega - {}^{\omega'}/_{2}\right) \cdot \widetilde{x^{*}}_{k}\left(\omega + {}^{\omega'}/_{2}\right)\right] = \int W_{ik}[\tau,\omega]d\tau \tag{7}$$

The tilde superscript in Equation (7) indicates Fourier-transformed time series. Because the Wigner-Ville distribution, unlike the wavelets, is sesquilinear rather than linear in amplitude of the signal, its integration over some domain is proportional to the cumulative energy concentrated in this time-frequency domain (Mallat, 1999).[2] Furthermore, a simple modification of Equation (5) allows to distribute a stochastic signal into uniform Heisenberg boxes:

$$\begin{gathered} \int \left(q - E(q)\right)^2 W(p,q)dpdq = \Delta q^2 \\ \int \left(p - E(p)\right)^2 W(p,q)dpdq = \Delta p^2 \\ \Delta p^2 \Delta q^2 \geq {}^{1}/_{4} \end{gathered} \tag{7a}$$

[2] For the definition of energy norm in stochastic context, see, e.g. Øksendal (2010).

Heisenberg-box analysis is a useful tool for distinguishing a relatively smooth signal from a relatively “spiked” noise (see Mallat, 1999). We plot the illustration to this thesis in Fig. 2.

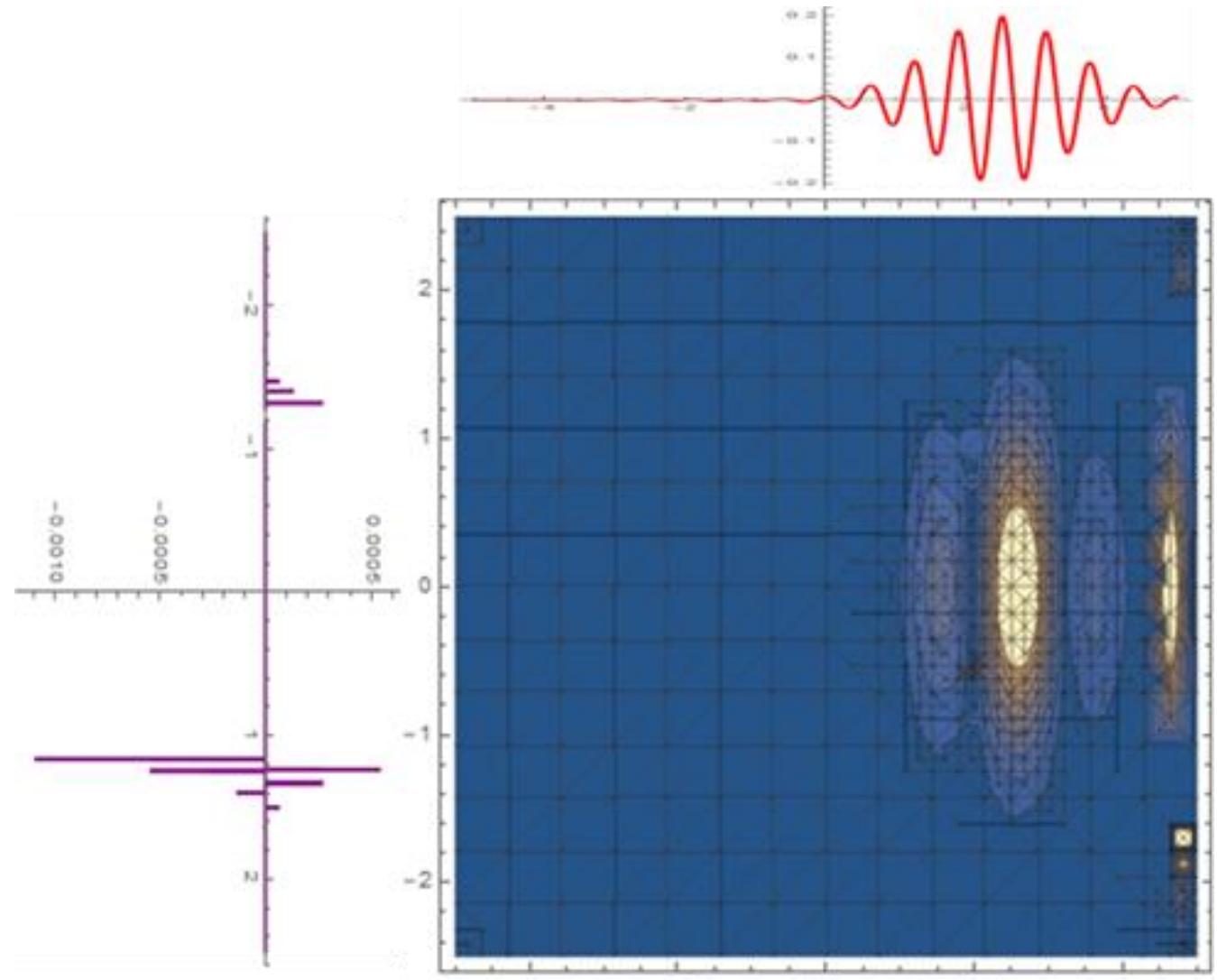

Fig. 2 Wigner function (square plot) of the two Gaussian-limited sinusoidal signals centered at t=-2 and t=2.5 separated by τ=4 with the amplitude ratio 1:100 (red curve). The presence of multiple frequencies is not clear from a time envelope but is very clear from the frequency spectra (purple curve).

There is an obvious (see, e.g. Chi and Russell, 2008) corollary of Equations (4) and (5), namely the “cross-correlation” property of the WVF. If the signal is a linear superposition of two signals,

$$x(t) = x_1(t) + x_2(t) \tag{8}$$

then,

$$W[\tau, \omega] = W_1[\tau, \omega] + W_2[\tau, \omega] + 2 \cdot Re[W_{12}[\tau, \omega]] \tag{9}$$

In Equation (11), the cross-correlation term $W_{12}$ is defined by the following Equation (compare to the Equation (4)):

$$W_{12}[\tau,\omega] = \int_{-\infty}^{\infty} x_1\left(\tau - {}^{t}/_{2}\right) \cdot x_2{}^{*}\left(\tau + {}^{t}/_{2}\right) e^{-i\omega t} dt \tag{10}$$

i.e. the off-diagonal terms of the matrix Wigner-Ville function.

Another important property of the Wigner function is that Wigner function of the normal distribution is the normal distribution of its arguments:

$$f(q) = e^{-\frac{q}{2\sigma^2}^{2}} \leftrightarrow W[p,q] = e^{-\frac{1}{\sigma^2}(p^2+q^2)} \tag{11}$$

In Equation (11), we neglect inessential normalization factors, which do not depend on the state space variables (coordinates).

The third important property is the explicit expression of the Wigner function of an arbitrary distribution in the form of an infinite series expansion over parabolic cylinder functions (Abramowitz and Stegun, 1965). An arbitrary distribution $f(q,t) \in L^2(\Omega)$ can be expanded into the series over the parabolic cylinder functions of integer order:
$U_n(q) = \left(\frac{\alpha^2}{4}\right)^{1/4} \frac{1}{2^{n/2}\pi^{n/2}} H_n(\alpha q) e^{-(\alpha q)^2/2}$ with α as a parameter. If the coefficients of this expansion are defined by the following equation:

$$f(q,t) = \sum_{n=0}^{\infty} c_n(t) \cdot U_n\,(q) \tag{12}$$

the Wigner function of this distribution can be expressed as:

$$W[p,q] = \sum_{n=0}^{\infty} |c_n(t)|^2\, V_n(p,q) \tag{13}$$

In Equation (8), the functions $V_n(p,q)$ are proportional to the Legendre functions of integer order (Abramowitz and Stegun, 1965):

$$V_n(p,q) = 2^n \sqrt{\pi}\, n!\, L_n(2(\alpha^2 q^2 + p^2)) \tag{14}$$

The above-mentioned fact of the connection between the Wigner function and the energy norm (Mallat, 1999) immediately follows from the series expansion of the Equations of (12) and (13).

Note, that the integrand of (10) represents an appropriately shifted correlation matrix. We shall use a cross-correlation terms implied by the WVF without special explanation as the measure of the interdependency of the time series $x_i(t)$ and $x_k(t)$. In our paper, we shall test interdependence of the residuals of the regression of the banks' idiosyncratic interest rate with LIBO

## 3 - Literature review

The suggested manipulation of the LIBOR by the group of banks participating in a survey caused such a tremendous shock for both the financial community and the general public because LIBOR is a frequent benchmark rate for a number of financial transactions: derivatives, mortgages, etc (*Economist*, 2012). Moreover, the LIBOR instruments are among the most liquid interest rate securities in their own right (Götsch, 2014). While the literature on the alleged LIBOR manipulation is enormous and of varying quality and accuracy, quantitative studies of the unseemly patterns (if any) in the behavior of the submitting banks are few.

Detailed analysis of the alleged manipulation of the LIBOR from the prospective of the Benford Law (Hill, 1995) as well as from the institutional standpoint, has been performed in the papers of Rosa Abrantes-Metz and collaborators (e.g. Abrantes-Metz, 2008 and 2010). J. Chen (2013) and Coulter and Shapiro (2015) proposed a quantitative model, yet, at present it is unclear how this model can be tested. Of existing studies modeling actual time series, I can point only to Eisl, Jankowitsch and Subramanyan (2014) and Snider and Youle (2012).

The word "manipulation" has many meanings. The definition closest to a textbook understanding was coined by Kyle and Viswanathan (2008) as any trading strategy which reduces price efficiency or market liquidity. Eisl, Jankowitsch and Subramanyan (2014) define manipulation to be any quote submission that differs from an honest and truthful answer to the question asked of the panel banks. Yet, these definitions are too vague to provide a significant quantitative guidance to the issuance of a manipulation verdict, though the tell-tale signals may be unmistakable.

In practice, both the papers by Snider and Youle (2012) and Eisl, Jankowitsch and Subramanyan (2014) propose sophisticated modes of quotes' distribution and view the outliers in the banks' submission of quotes in the model-prescribed distribution of LIBOR as evidence for manipulation. The analysis provided in these papers is quite convincing, yet it does not exclude the possibility that common macroeconomic factors were the culprit of clustering behavior. In particular, Eisl, Jankowitsch and Subramanyan consider quote submissions as sampling of a stable rate distribution unique for all members of fellowship. This approach underlies a much richer structure than it is possible in the current paper, yet it ignores the possibility that the difference in quotes resulted from different macroeconomic conditions of the domicile nations as well as the credit situation of the member banks. This paper uses a model-free approach of clustered correlations in phase-space between the pairs of individual banks on a common LIBOR quote.

## 4 - Stylized Model of Interbank Borrowing with the Participating Authority of a Central Bank

In this section, we describe a stylized model of interbank borrowing. Namely, we are considering two agents: a representative lender (supplier of funds) and a representative borrower (demander of funds). Because of the existence of reserves, modern banks can borrow at a rate approximating risk-free rate up to a certain quantity Q*. After that, the admissible rate rises with the amount borrowed and it benefits the bank to switch from lending to borrowing. The lender's situation is the opposite. It can supply funds at a risk-free rate (rate of borrowing from the Central Bank) plus a certain spread to justify lending activity up to a certain quantity Q**. If the supply of funds exceeds Q**, it has to offer discount to this rate to place available funds (Fig.3A), or switch from borrowing to lending.

Nothing in this stylized picture indicates the existence of an equilibrium, or clearing, in the parlance of banking. However, the Central Bank is tasked with maintaining such equilibrium, so it can supply or withdraw funds (by buying/selling T-bonds in requisite quantities), until the equilibrium in credit market is restored (Fig. 3B). However, the capacity of the Central Bank to supply additional funds to a member bank is limited by the regulatory constraints. Outside these constraints, a private bank must borrow new funds on the open market, or switch from borrowing to lending. At the equilibrium, the bank should be indifferent with respect to its position as a net borrower or a net lender (Fig. 3C). The bank's offer rate must be equal to the market ask rate with a spread, reflecting residual market imbalances.

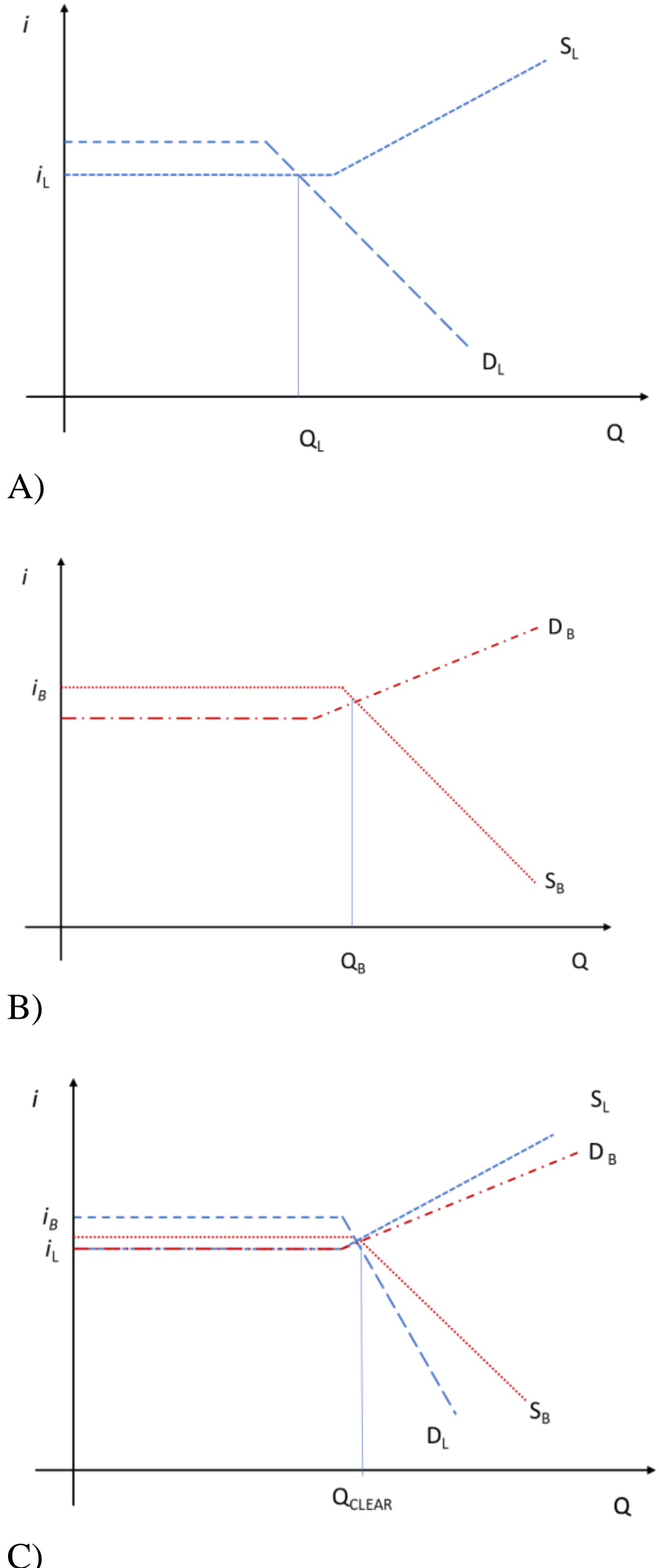

Fig. 3 A) Stylized supply-demand curve for the interbank borrowing from the lender's prospective. The lender bank can supply funds to the borrower at a small spread with a Central Bank rate up to exhaustion of reserve requirements. Above that quantity, it benefits the bank to switch from lending to borrowing funds. B) Stylized supply-demand curve from the borrower's prospective. The borrower can borrow funds at a small spread with a Central Bank up to its credit-rating defined credit limits. Above that quantity, it benefits the bank to switch from borrowing to lending. C) Interbank market clears at the inflection point when both borrower and lender are indifferent to switching their roles.

## 5 - Description of the Methodology

The institution in charge of setting the LIBOR during the analyzed period was the British Bankers Association (BBA). The mechanism of the submission of the quotes by the "fellowship of the LIBOR" was such: eighteen banks, which were presumably the largest players in the London Interbank Currency Market, submitted daily quotes of the interest rates by which they borrow money (or think other banks can borrow them—in case of the EURIBOR). The four lowest and the four highest quotes were then omitted as to diminish the influence of the outliers and to prevent outright manipulation. The authority for setting the LIBOR has been now transferred to the Intercontinental Exchange (ICE), published by Thomson Reuters and is regulated by the British Financial Services Authority (Whitley, 2011, Economist, 2012, Wikilbor, 2015).

Certainly, quotes, which are persistently in the lower or upper band of rates submissions can indicate a particularly high or low creditworthiness of the individual banks rather than their bias in estimating the rates. In this paper, I apply the following method to identify possible patterns of non-random behavior, or clustering, which can be viewed as instances of manipulation. First, I try to de-trend the individual LIBOR quotes by their regression with the quoted LIBOR rate. For the six banks from the entire 18-bank sample, I also condition them on the credit quality measured by regression of the yield on their CDS with the interest rate of their domicile.

The residuals of the de-trended regressions were then analyzed using the Wigner-Ville function (see the Section 7). My toy example of the LIBOR time series filtering is being used in the current paper to advertise a new and powerful method, which can be applied in other fields of finance.

These quotes were ranked in the order of magnitude, then the rate setter omitted the four lowest and the four highest quotes and then provides the unweighted average of the rest of the submitters. In Table 1, I provide the list of the banks in the sample together with their domicile. In the Fig. 4, the daily quotes of select banks and the quoted LIBOR taken from Bloomberg© are being presented for the period from 4/17/2011 to 7/17/2012 (313 business days).

**Table 1** Banks participating in the submission of the LIBOR data in the sample dated from 4/17/2011 to 7/17/2012.

| *No.* | *Bank* | *Domicile* |
|---|---|---|
| 1. | Barclays | UK |
| 2. | JPM Chase | US |
| 3. | BTMU | Japan |
| 4. | BOFA | US |
| 5. | BNP Paribas | France |
| 6. | CA-CIB | France |
| 7. | Citibank | US |
| 8. | Credit Suisse | Switzerland |
| 9. | Deutsche Bank | Germany |
| 10. | HSBC | UK |
| 11. | Lloyds | UK |
| 12. | Norinchukin | Japan |
| 13. | Rabobank | Holland |
| 14. | RBC | Canada |
| 15. | RBS | UK |
| 16. | Societe Gen | France |
| 17. | Sumitomo | Japan |
| 18. | UBS AG | Switzerland |

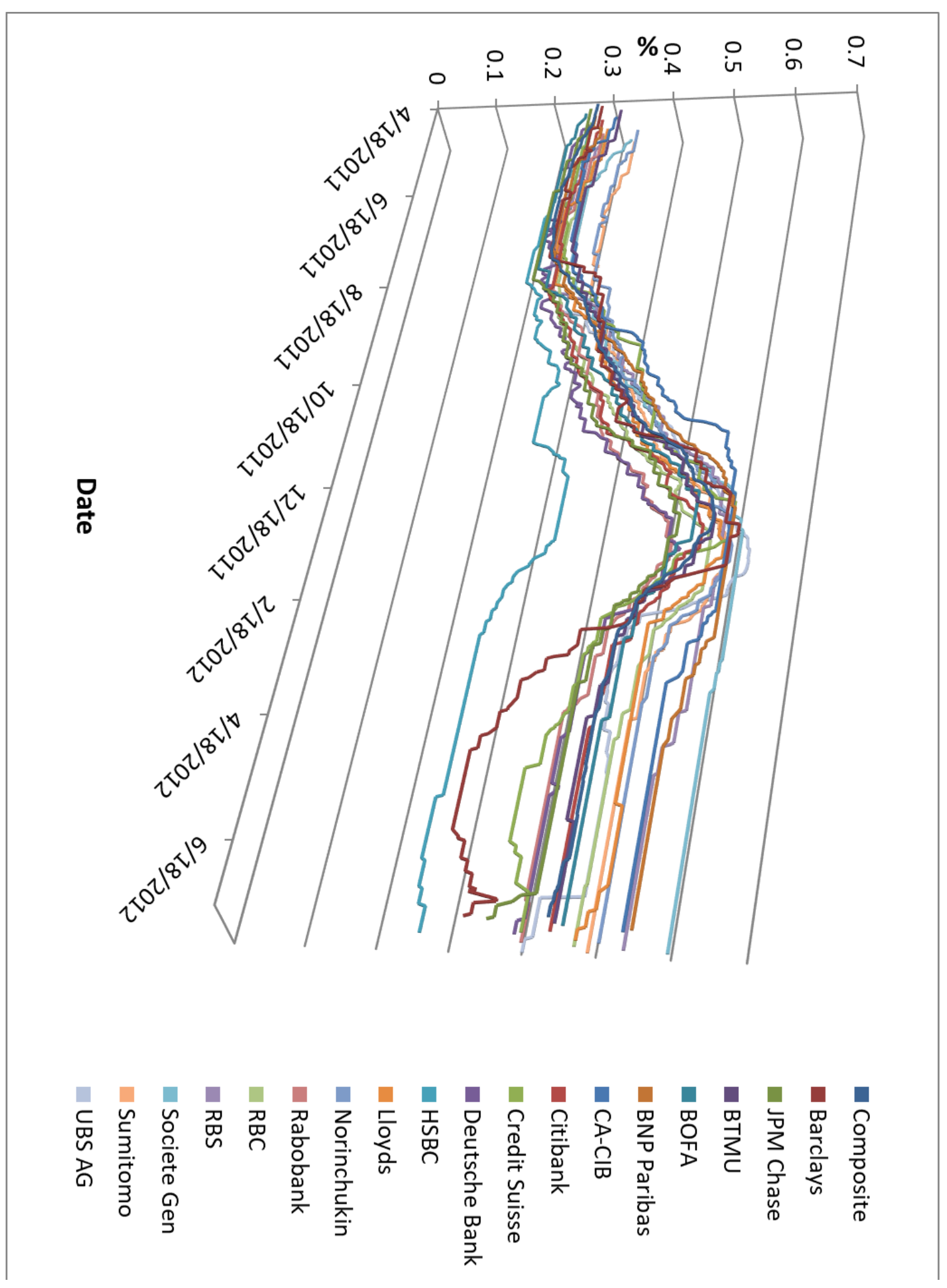

Fig. 4 The plot of the LIBOR dynamics and submission by 18 separate banks during the 15 months' period between 4/17/2011 and 7/17/2012.

To de-trend the banks submissions, I ran an OLS daily regression with the LIBOR

$$r_{it} = \alpha_i + \beta_i \cdot r_{LIBOR,t} + e_{it} \tag{15}$$

In Table 2, the results of the linear regression of Equation (15) of the panel of banks enumerated in the order of the Table 2 are being presented. All individual β's in the panel except HSBC (No. 10) are close to each other and to unity but also are statistically distinguishable from β=1, at least, with a 5% margin of error.

The HSBC data in the Table 2 are clearly the outliers in the whole dataset. This can indicate that its tight connection with Chinese market imparts a feature of a partially controllable currency, the renminbi on the lending and borrowing practices of this bank. Barclays of UK had its own problems before the sample period (Snyder and Youle, 2012) and will be further omitted from the sample.

**Table 2** Extensive OLS statistics according to Equation (15) for the Fellowship of the LIBOR in the period 4/17/2011-7/17/2012 (313 business days). The order of banks corresponds to Table 1. Row 19 indicates LIBOR itself. Rate variances and standard deviations of α and β are given in parentheses. For most of the banks (except Barklays and HSBC) the correlation with the LIBOR exceeded 90%. All β's except the one for HSBC are numerically close to one but the difference from unity is statistically significant at 5% accuracy.

| **Bank No.** | **Rate, %** | **α** | **β** | **$R^2$** |
|---|---|---|---|---|
| **1** | 0.4103 (0.0117) | 0.0382 (0.0147) | 0.8287 (0.0346) | 0.647 |
| **2** | 0.3783 (0.0123) | -0.0042 (0.0030) | 0.9196 (0.0070) | 0.982 |
| **3** | 0.3731 (0.0100) | 0.0678 (0.0019) | 0.8486 (0.0046) | 0.991 |
| **4** | 0.4160 | -0.0321 | 1.0467 | 0.985 |
| | (0.0085) | (0.0031) | (0.0073) | |
| **5** | 0.3973 | -0.0645 | 1.2759 | 0.952 |
| | (0.0130) | (0.0069) | (0.0163) | |
| **6** | 0.4590 | -0.0107 | 1.1660 | 0.963 |
| | (0.0199) | (0.0055) | (0.0130) | |
| **7** | 0.4677 | 0.0055 | 0.9418 | 0.978 |
| | (0.0165) | (0.0034) | (0.0080) | |
| **8** | 0.3919 | 0.0498 | 0.8391 | 0.827 |
| | (0.0106) | (0.0092) | (0.0217) | |
| **9** | 0.3941 | -0.0016 | 0.8813 | 0.980 |
| | (0.0099) | (0.0030) | (0.0071) | |
| **10** | 0.3601 | 0.1528 | 0.2691 | 0.643 |
| | (0.0092) | (0.0048) | (0.0113) | |
| **11** | 0.2632 | -0.0107 | 1.0551 | 0.983 |
| | (0.0013) | (0.0033) | (0.0079) | |
| **12** | 0.4222 | 0.0731 | 0.9228 | 0.993 |
| | (0.0132) | (0.0019) | (0.0044) | |
| **13** | 0.4517 | 0.0497 | 0.7736 | 0.991 |
| | (0.0100) | (0.0018) | (0.0041) | |
| **14** | 0.3672 | -0.0140 | 1.0252 | 0.976 |
| | (0.0070) | (0.0038) | (0.0090) | |
| **15** | 0.4067 | -0.0534 | 1.2253 | 0.953 |
| | (0.0125) | (0.0065) | (0.0154) | |
| **16** | 0.4493 | -0.0435 | 1.2649 | 0.921 |
| | (0.0184) | (0.0089) | (0.0210) | |
| **17** | 0.4755 | 0.0728 | 0.8927 | 0.996 |
| | (0.0202) | (0.0014) | (0.0032) | |
| **18** | 0.4391 | -0.0406 | 1.0990 | 0.943 |
| | (0.0093) | (0.0065) | (0.0152) | |
| **19** | 0.4103 | 0.0000 | 1.0000 | 1.000 |
| | (0.0149) | (0.0000) | (0.0000) | |

## 6 - Controls for the Credit Quality of Select Banks

As was already mentioned in the Introduction, large systematic deviations from the LIBOR might result not only from an abusive manipulation of the LIBOR quotes but from the differences in the credit quality of the submitting banks. Obviously, a bank which has submissions repeatedly below LIBOR can increase its rates to be more in line with the average. Snider and Youle (2012) and Eisl, Jankowitsch and Subrahmanyan (2014) who used actual quotes did not control for this effect. This paper does not fully exclude its possibility either, but we assume that the CDS issued on their credit portfolios of the member banks should eliminate this inefficiency.

The credit quality of the banks can be measured directly through CDS. (Das, 2005) However, most individual obligor CDS are not very liquid and cannot be expected to reflect their credit situation on a daily basis. That’s why we use correlation of the CDS yields with the national benchmark interest rate (for instance, yield on 3 month Treasury notes for the US), to infer daily fluctuations of the credit quality of the banks. [3]

Technically, we should have been using the CDS of the creditors, not the issuing banks. This would be impractical because each bank in question issues credit to a large number of borrowers including fellow members of LIBOR community. However, due to the stylized model of the Section 4, in the clearing equilibrium, the LIBOR rate should be close to the rate at which a given bank is indifferent to switching from being lender to being a borrower in the LIBOR market.

This procedure omits idiosyncratic errors due to the operational changes in a given obligor because those can also indicate abusive behavior. The CDS statistic according to Bloomberg© in the period 01/01/2012-07/17/2012 is sufficient to produce correlations with their national short rate for five banks, i.e. one-third of the entire sample. The data on CDS spreads of six banks are listed in Table 3A. Table 3B lists the results of the detrending of the LIBOR quotes of the six banks with the estimated CDS rate as an additional factor:

[3] The betas for the CDS were used by Tang and Yan in the context of credit default swap illiquidity (Tang, 2009) and by Bai and Collin-Dufresne in the context of CDS-bond basis spread (2013). Bai and Collin-Dufresne also used daily LIBOR as an explanatory factor for this basis. The data were available to author from Bloomberg© only for these six banks.

$$\begin{aligned} r_{it} &= \alpha_i + \tilde{\beta}_i \cdot r_{LIBOR,t} + \theta_i \cdot \hat{r}_{CDSi,t} + \varepsilon_{it} \\ \hat{r}_{CDS,it} &= \alpha_{credit,i} + \beta_{credit,i} \cdot r_{riskfree,t} + v_{it} \end{aligned} \quad (16)$$

where *i=2, 4, 5, 6, 7* and *9*. In Equation (16), $\hat{r}_{CDSi,t}$ is an estimated credit quality of a given bank in the sample according to Table 3A. Tilde over β factor, which has the same meaning as the simple credit beta in Equation (15), is a reminder that a correlation with the LIBOR can change very significantly because of the potential collinearity of the national short rates with the LIBOR.

The residuals $\varepsilon_{it}$ of the regression of the Equation (3) will be used as the primary time series in subsequent correlation studies. While the numerical value of the LIBOR quotes and their deviation from the mean of different participating banks can be similar because of a similar credit situation, residuals should pose no such problem.

Before we demonstrate our method, we provide an additional robustness check. Namely, we ran a version of the white-noise test on the residuals, recently proposed by Bagchi, Characiejus and Dette (2017). This test is based on estimation of the quantity $\xi = \sqrt{\pi T}\,\frac{\widehat{M}_T^2}{4\,\hat{\sigma}_T}$ where $T$—the length of the time series, $\widehat{M}_T^2$ is an estimator of the sum of squares of covariance matrix and $\hat{\sigma}_T$ is an estimator of squares of the residuals. Test results are provided in Table 3C. We can reject the null hypothesis that our residuals are white noise for all our time series, except for Credit Agricole, but in this test we cannot possibly determine the dates of regime change or an existence of data clustering.

**Table 3A** Correlation of the Credit Spread with the National Short Rate.
We present the results of regression of the second Equation (3) predicting the daily CDS spreads on a single obligor (bank) from the national short rates. The value of $\alpha_{credit}$ and $\beta_{credit}$ are inferred from a regression of CDS spread for select banks with the national short rate (3 Month T-bill, in case of USA, 3-Month note of Banque de France and T-note of the Bundesbank) taken from 01/01/2012 to 07/17/2012. The regression gives prediction of CDS rate in basis points (bp) as a function of short rate in percentages.

| *Bank* | $\alpha_{credit}$, *bp* | $\beta_{credit}$, *bp/%* |
|---|---|---|
| BOFA | 506.93 | -29.47 |
| Citibank | 392.68 | -5.03 |
| JPMorgan | 295.09 | -4.43 |
| Credit Agricole | 543.17 | -66.68 |
| Deutsche Bank | 339.31 | -5.23 |
| Societé Generale | 539.21 | -11.81 |

**Table 3B** Regression of Equation (3) for Select Banks.
Table represents the OLS regressions with credit quality of a given bank as the control variable. Variances for specific coefficients are given in parentheses.

| *Bank* | $\alpha$ | $\beta$ | $\theta$ | $R^2$ |
|---|---|---|---|---|
| BOFA | 0.0124<br>(0.0035) | 1.0243<br>(0.0054) | -0.0098<br>(0.0006) | 0.992 |
| Citibank | -1.2625<br>(0.0610) | -0.2496<br>(0.0248) | 0.4778<br>(0.0157) | 0.801 |
| JPMorgan | -0.9987<br>(0.0519) | -0.2175<br>(0.0247) | 0.5350<br>(0.0177) | 0.793 |
| Credit Agricole | -0.4285<br>(0.0459) | 0.9499<br>(0.0263) | 0.0985<br>(0.0108) | 0.970 |
| Deutsche Bank | 11.8723<br>(1.3497) | -3.5003<br>(0.3979) | 0.8671<br>(0.0065) | 0.984 |
| Societé Generale | -7.485<br>(0.3875) | 0.7070<br>(0.0323) | 1.4356<br>(0.0748) | 0.964 |

**Table 3C**. White noise test of covariance matrices due to Bagchi, Characiejus and Dette (2017). Auto-covariance matrices of the five banks were tested for the statistic $\xi = \sqrt{\pi T}\ \frac{\widehat{M}_T^2}{4\ \hat{\sigma}_T}$ where $T$=313, $\widehat{M}_T^2$ is a sum-of-squares estimator for the covariance matrix and $\hat{\sigma}_T$ is an estimator of squares of the residuals. The only covariance matrices exhibiting no difference from the white noise at 5% are those with Credit Agricole.

| *Pair* | *ξ* | *Pr(x>ξ)* |
|---|---|---|
| Citi-JPM | 2.49 | 1.8% |
| Citi-Credit Agricole | 2.05 | 4.9% |
| Citi-Deutsche | 2.61 | 1.3% |
| JPM-Credit Agricole | 1.97 | 5.7% |
| JPM-Deutsche | 2.42 | 2.1% |
| Credit Agricole-Deutsche | 1.95 | 6.0% |

## 7 - Wigner-Ville function analysis

The essence of our statistical method to examine residuals of regressions (2) and (3) in more detail is to calculate correlation of the aliased maps/arrays (Fig. 5) of the covariance function between the residuals of the panels of regression of Equation (3). The method of aliasing was borrowed from image recognition methodology (Mallat, 1999).

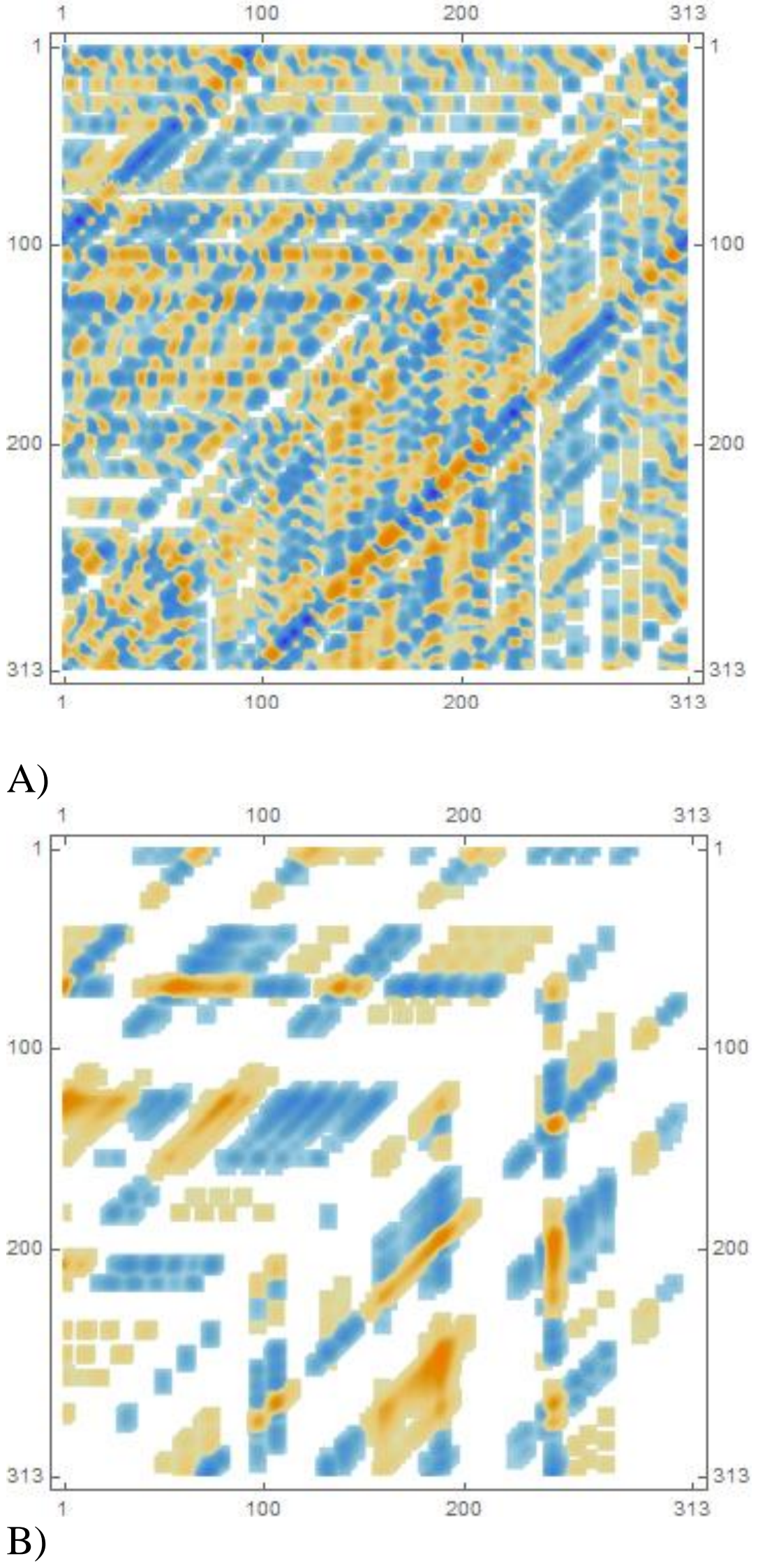

A)

B)

Fig. 5 A) The autocorrelation map of the aliased simulated random walk is almost ergodic. Orange points indicate positive values and blue points indicate the negative values. Data points are subjected to Gaussian blur. B) The autocorrelation map of the typical aliased residuals of the regression of Equation (3) with the same Gaussian blur as in Fig. 5A. Clustering of the extreme events is clearly discernible.

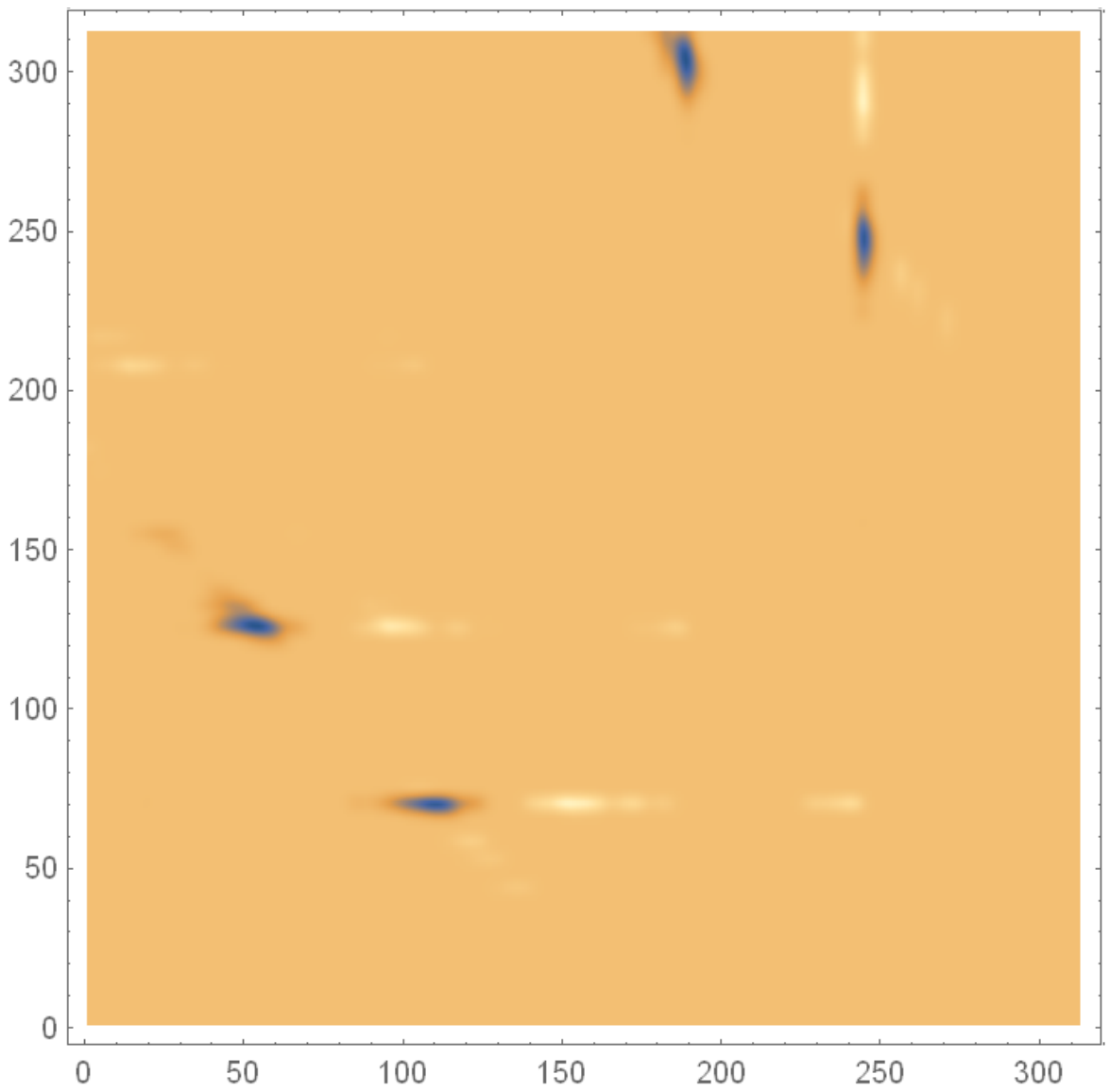

Fig. 6 Densitogram of the autocorrelation map for the pair of banks with low correlation of Wigner matrices (JP Morgan-Paribas). The difference between density of clusters here and in Fig. 5B is clearly visible.

Matrix arrays were aliased to exclude a random noise in data as to assert meaningful correlations on the level of 36 standard deviations for the entire array, i.e. roughly at $36/\sqrt{313\ days}$ =2.03 standard deviations for a single element/event. Obviously, an application of square root rule typical for the normal distributions to the array of unknown statistics is spurious but it is provided as a guide. The threshold of 36 $\sigma_1$=2.03$\sigma_N$ was chosen because for a higher threshold there are too few events to generate credible statistics and for a lower threshold the probability of the random event exceeding it is too high. Naïve probability for so defined two-tailed event is P(x>2.03)=5.1%. Similar 5% threshold for a comparable number of events (150-250) has been used by Snyder and Youle.[4] However, there is no basis for assuming normal distribution of these events.

The resulting WVF is the Fourier transform of the specially prepared covariance matrix (see Equation (1)). A typical plot of a modulus of WVF is shown in Fig. 7. An obvious intuition is that the residuals of a bank's submissions must not have any distinguishable pattern if their LIBOR submissions were accurate and independent, but should demonstrate a distinct pattern in case of collusion and/or falsified submission.

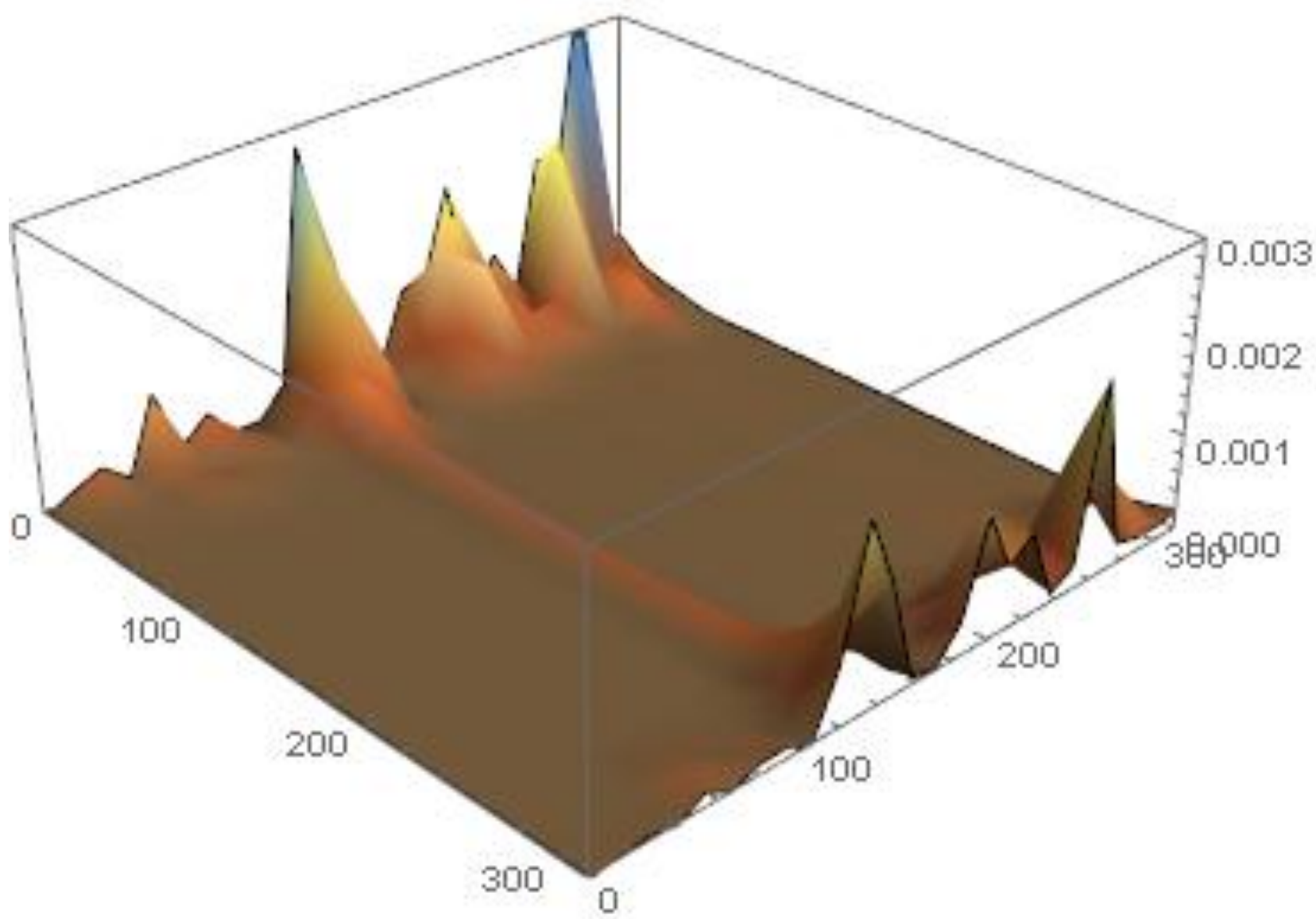

Fig. 7 The plot of the absolute value of the WVF of the residuals of a typical financial time series from our "fellowship" sample. We identify the ridges with the regime changes possibly indicating manipulative behavior.

These maps represent not only the temporal but also the frequency variation of the pairs of residuals of the bank quotes. The author is agnostic at this point, whether exceptionally high or unusually low correlation is indicative of market manipulation by the banks.

The correlations are assembled in Table 4.

[4] Snyder and Youle (2012), p. 13.

**Table 4** Correlations of the Wigner-Ville matrix arrays of regression residuals between select banks.
"City" is Citibank, "JPM"=J. P. Morgan, "CA"=Credit Agricole, "DB"=Deutsche Bank and "BNP"=Banc Nacionale Paribas. Paribas data are given in grey to indicate that its residuals were not controlled for the credit quality of the bank.

| | *City* | *JPM* | *CA* | *DB* | *BNP* |
|---|---|---|---|---|---|
| City | 1 | 0.981 | 0.142 | 0.515 | -0.036 |
| JPM | | 1 | 0.166 | 0.360 | 0.051 |
| CA | | | 1 | 0.435 | 0.497 |
| DB | | | | 1 | 0.415 |

We observe an almost perfect correlation of the residuals of the Citibank and J. P. Morgan, correlations of the French banks among themselves are moderate to low—with respect to other banks and that Deutsche Bank has moderate correlation among both American and French banks. So far, no quantitative criterion has been established by econometrics to distinguish between "high" or "low" correlation, which is still in the eye of the beholder. We must mention though that, for our two-dimensional maps, a high degree of correlation of Wigner functions is equivalent to clustering in phase space (see below). Unless as yet hidden parameter exhibits clustering on its own, this conclusion suggests conscious manipulation.

The author tested the same algorithm on: (1) a set of normally distributed random numbers—corresponding to the random walk hypothesis for the submitted rates—and (2) the white noise distributed as differences of Lévy flights (Paul and Baschnagel, 2006). The random walks for the same number of events (313×313) typically exhibit correlations in the range of 0-20%. A typical aliased covariance matrix of these tests is demonstrated in Fig. 5A.

The most glaring difference between the random distributions and the Wigner functions of the "suspect" banks is that maps of random numbers do not demonstrate any clustering between days and close frequency bins. If there is no manipulation, one would expect that the residuals of submitted quotes should be as random as a set of random numbers but there can be caveats. The author is currently working at establishing the quantitative criteria for daily and frequency-wide length/duration of such clustering.

## 8 - Conclusion

We used analysis of the patterns in the vector Wigner-Ville Function to test correlation between de-trended LIBOR quotes of the select "LIBOR fellowship" banks (Citibank, J. P. Morgan, Deutsche Bank and Credit Agricole) in both the temporal and the frequency domains. A study of the WVF is proposed as the new method to identify patterns in the financial time series.

The submission quotes time series were controlled for the credit quality of the obligor and the national interest rate (Equation 3). Covariance matrix of residuals was filtered so that only tail events exceeding 2.03 standard deviations for the element were included in the tally. A naïve probability of the event being included in our tally is 5%, which would provide ~50,000 analyzable events in a sparse matrix with 313×313 elements.

Our analysis established an almost 100% correlation of the tail events for the submission of quotes by Citibank and J. P. Morgan. This correlation is viewed by us as a possible indicator of manipulation of the LIBOR quote submission process by one of the banks or both. Clustering of unusual events was detected by a ridge detection algorithm and indicated that these unusual events happened at the end of July, 2011, in mid-October, 2011 and in mid-February 2012 (Figs. 7 and 8). Interestingly, the ridges identified by a detection algorithm on a two-dimensional map are relatively robust with respect to the pair of banks one chooses for the identification.

Our method of the Wigner-Ville function thus allows us not only to reject the randomness of quote deviations but, at variance with the standard white noise tests (Table 3C) it can tentatively establish when the alleged cheating took place. While WVF analysis is only qualitative like all image recognition, heuristic identification of the data stamp of cheating can significantly simplify the job of the investigator (Zitzewitz, 2012).

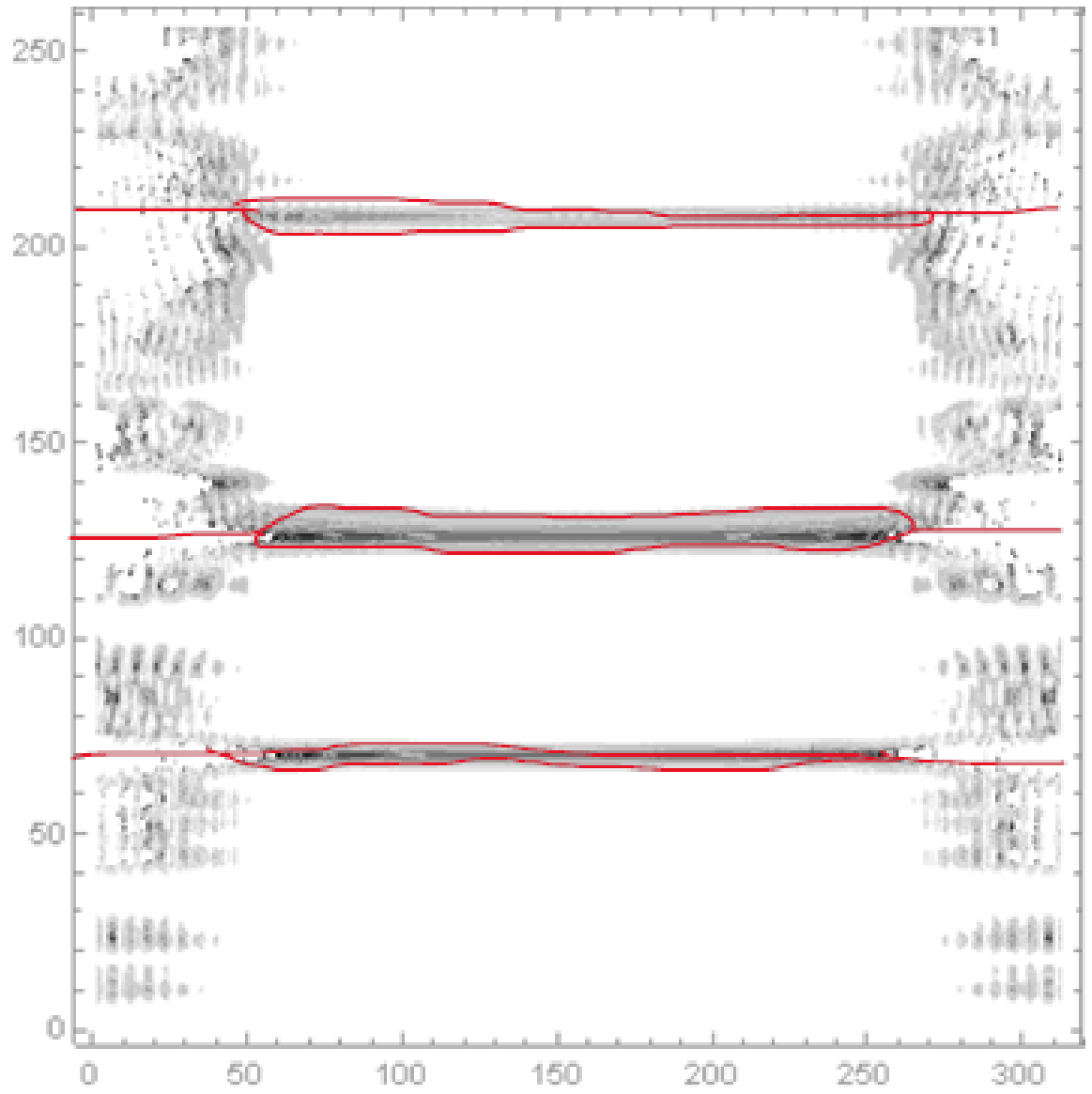

Fig. 8 The ridges in the plot of the Wigner function (see Fig. 7) identified by a ridge detection algorithm (highlighted in red). According to them, we can identify and date regime changes in the time series as happening around business days 70, 125 and 210 in the sample. The position ridges are relatively robust with respect to the pairs of banks with low and high correlation of Wigner matrices.

The case with Deutsche Bank, Credit Agricole, Sociéte Generale and Paribas (the two last banks were used for partial calibration) is much more complex. At a glance from Table 4, one can assume that a correlation of Wigner-Ville arrays at 50% indicates just a domicile effect not related to the domestic interest rates or credit quality of a selected bank. Then, our analysis did not establish enhanced correlations between the French banks (Credit Agricole and Paribas). Deutsche Bank can fit this recognition pattern if one assumes that 50% of its LIBOR quoting was tied to the American and 50%—to the Continental European events. Alternative explanation would be that 50% of the time its traders abandoned connection with its domicile to copy-cat American banks.

In 2011, the *Wall Street Journal* reported that Bank of America, Citigroup and UBS AG were under investigation for manipulation of

quotes—the first two being a part of our sample (WSJ, 2011). Currently, the judgment of manipulation of the LIBOR quotes has been issued against Barklays Bank by the US CFTC ($200 m.), US Department of Justice ($160 m.) and UK Financial Services Authority—Lb. 59.5 m. (US Department of Justice, 2012, Eisl, Jankowitsch and Subramanyan, 2014). Deutsche Bank was fined for unprecedented $2.5 billion in 2015. So far, no American bank was held responsible by the US authorities (Reuters, 2015). This paper does not delve into the politics of the US regulators, Department of Justice and courts.

## Appendix A. **Dynamic Equation for the Wigner Function**

The dynamic evolution equation for the Wigner-Ville function is complicated, which probably, slowed down its acceptance by the econometric community. Yet, there is no comparable dynamic equation for the wavelets, which did not influence their wide acceptance. For the conventional definition of the Wigner-Ville function, the evolution equation for the Wick-tilted ($\tau \to it$) dynamical equation (A.1) is as follows:

$$i\frac{\partial\psi}{\partial t}+\frac{1}{2}\frac{\partial^2\psi}{\partial x^2}+V(x)\psi \tag{A.1}$$

$$\frac{\partial W[p,x,t]}{\partial t}+p\frac{\partial}{\partial x}W[p,x,t]=-2\pi i\int\left[V\left(x+\frac{y}{2}\right)-V\left(x-\frac{y}{2}\right)\right]\times \psi\left(x+\frac{y}{2}\right)\psi^*\left(x-\frac{y}{2}\right)e^{-ipy}dy \tag{A.2}$$

Equation (A.2) is not only an integro-differential equation but also, as a form of Liouville equation, has difficult computational properties. There is no similar equation for the Ville function of Equation (3).

## Appendix B. **Modified definition of the Wigner function for the mathematical finance**

One of the main shortcomings of the Wigner function approach is the complexity of the explicit closed-form equation for the Wigner function dynamics for all but the simplest random processes (Appendix A and Hillery, O'Connell, Scully and Wigner 1984). However, one can deduce an approximate dynamical equation, which can be satisfactory in many cases.

Here, we provide a slight modification of the Wigner function, which is adapted to the formulation of problems as they exist in mathematical finance. The difference in formulations results mainly from the fact that WVF was designed with the Schrodinger equation and Feynman measure in mind, while mathematical finance deals with diffusion-type equations and the Wiener measure. Namely, if we have a generator $L$ for a random process (e.g. Borodin and Salminen, 2005), we can write a forward evolution equation in the form:

$$\frac{\partial f(q,t)}{\partial t} - Lf(q,t) = 0 \qquad \text{(B.1)}$$

where $q$, as before is the vector of the state variables, and the backward evolution equation in the form:

$$\frac{\partial \tilde{f}(q,t)}{\partial t} + L^{*}\tilde{f}(q,t) = 0 \qquad \text{(B.2)}$$

Then, we define Wigner function as the Laplace transform of the following bilinear combination:

$$W[p,q,t] = \int_0^{\infty} f(q + {q'}/_{2}, t) \cdot \tilde{f}(q - {q'}/_{2}, t) \cdot e^{-p \cdot q'} dq' \qquad \text{(B.3)}$$

The integration in Equation (B.3) is conducted over the space of the state variables. If the operator L is diffusion:

$$L = a(x)\frac{\partial^2}{\partial x^2} + b(x)\frac{\partial}{\partial x} + c(x) \qquad \text{(B.4)}$$

(We assume the state space as 1-dimensional and denote as *x* the only state variable; the generalization for the Euclidian vector state space is straightforward), one can obtain an <u>approximate</u> dynamic evolution equation for the Wigner function in the limit of slow dependence of the Wigner function on its arguments. In that approximation, the integrand of the true dynamic equation (see Hillery, O'Connell, Scully and Wigner, 1984) can be replaced with its Taylor series, which we truncate up to the second order in derivatives.

This equation has the form:

$$\partial W[p,x,t]/_{\partial t} = a(x)\frac{\partial^2 W}{\partial x^2} - \frac{\partial b(x)}{\partial x} \cdot \frac{\partial W}{\partial p} + cW + \frac{1}{2}c''(x)\frac{\partial^2 W}{\partial p^2} \qquad \text{(B.5)}$$

In Equation (B.5) and below, the dot means the time derivative and the double apostrophe—the x-derivative. Note, that if *b* and *c* in the definition of diffusion of Equation (B.4) are constants, *p* and *x* variables in the Equation (B.5) can be separated:

$$W[p,x,t] = \bar{W}[x,t] \cdot \bar{\bar{W}}[p] \qquad \text{(B.6)}$$

The Equation (B.5) then obtains a familiar diffusion form for the rates (Black-Scholes equation for prices):

$$\dot{\bar{W}} = a(x)\frac{\partial^2 \bar{W}}{\partial x^2} + c\bar{W} \qquad \text{(B.8)}$$

As is conventional in this context, the coefficient *a(x)* is identified with half of the volatility squared and *c*—with the risk-free rate. The function $\bar{\bar{W}}[p]$ can be determined from the boundary conditions.